\documentclass[twocolumn]{aastex63}
\usepackage{ amsmath}

\defcitealias{vanVelzen+21}{V21}

\newcommand{\be}{\begin{equation}}
\newcommand{\ee}{\end{equation}}

\shorttitle{Emission from Cooling TDE Envelopes}
\shortauthors{B.D.~Metzger}
\graphicspath{{./}{figures/}}

\begin{document}

\title{\Large{Cooling Envelope Model for Tidal Disruption Events}}


\author[0000-0002-4670-7509]{Brian D.~Metzger}
\affil{Department of Physics and Columbia Astrophysics Laboratory, Columbia University, New York, NY 10027, USA}
\affil{Center for Computational Astrophysics, Flatiron Institute, 162 5th Ave, New York, NY 10010, USA} 

\begin{abstract}
We present a toy model for the thermal optical/UV/X-ray emission from tidal disruption events (TDE).  Motivated by recent hydrodynamical simulations, we assume the debris streams promptly and rapidly circularize (on the orbital period of the most tightly bound debris), generating a hot quasi-spherical pressure-supported envelope of radius $R_{\rm v} \sim 10^{14}$ cm (photosphere radius $\sim 10^{15}$ cm) surrounding the supermassive black hole (SMBH).  As the envelope cools radiatively, it undergoes Kelvin-Helmholtz contraction $R_{\rm v} \propto t^{-1}$, its temperature rising $T_{\rm eff} \propto t^{1/2}$ while its total luminosity remains roughly constant; the optical luminosity decays as $\nu L_{\nu} \propto R_{\rm v}^{2}T_{\rm eff} \propto t^{-3/2}$. Despite this similarity to the mass fall-back rate $\dot{M}_{\rm fb} \propto t^{-5/3}$, envelope heating from fall-back accretion is sub-dominant compared to the envelope cooling luminosity except near optical peak (where they are comparable).  Envelope contraction can be delayed by energy injection from accretion from the inner envelope onto the SMBH in a regulated manner, leading to a late-time flattening of the optical/X-ray light curves, similar to those observed in some TDEs.  Eventually, as the envelope contracts to near the circularization radius, the SMBH accretion rate rises to its maximum, in tandem with the decreasing optical luminosity.  This cooling-induced (rather than circularization-induced) delay of up to several hundred days, may account for the delayed onset of thermal X-rays, late-time radio flares, and high-energy neutrino generation, observed in some TDEs.  We compare the model predictions to recent TDE light curve correlation studies, finding agreement as well as points of tension.

\end{abstract}


\section{Introduction} 
\label{sec:intro}

A tidal disruption event (TDE) occurs when a star orbiting a supermassive black hole (SMBH) on what is typically a parabolic orbit, comes sufficiently close to the SMBH to be strongly compressed and torn apart by tidal forces \citep{Hills75,Luminet&Carter86,Rees88,Evans&Kochanek89,Stone+13,Guillochon&RamirezRuiz13,Coughlin&Nixon22}. 

When TDE flares were first discovered in UV (e.g., \citealt{Stern+04,Gezari+06,Gezari+08}) and optical (e.g., \citealt{vanVelzen+11,Cenko+12,Arcavi+14,Chornock+14}) surveys, one largely (but not wholly; \citealt{Loeb&Ulmer97}) unexpected discovery were their high optical luminosities $L_{\rm opt} \gtrsim 10^{43}$ erg s$^{-1}$, modest effective temperatures $T_{\rm eff} \approx 10^{4.2}-10^{4.7}$ K, and correspondingly large photosphere radii $\approx 10^{14}-10^{15}$ cm (e.g., \citealt{Arcavi+14,Holoien+14,Hung+17,vanVelzen+21}; see \citealt{vanVelzen+20}, \citealt{Gezari21} for recent reviews).  It was previously a common assumption that once the tidal streams of the disrupted star dissipate their bulk kinetic energy (``circularize''), the resulting structure would be a compact disk orbiting the SMBH comparable in size to the tidal sphere (typically tens or hundreds of gravitational radii, or $\sim 10^{13}$ cm for $10^{6}-10^{7}M_{\odot}$ SMBH), which produces multi-color blackbody emission peaking in the soft X-ray bands (e.g., \citealt{Rees88,Cannizzo+90,Ulmer99,Lodato&Rossi11}) with only a tiny fraction of the luminosity radiated at optical/UV frequencies.

After reaching peak luminosity, many TDE optical light curves decay roughly following a $\propto t^{-5/3}$ power-law (e.g., \citealt{Gezari+06,Hung+17}), thus appearing to track the rate of mass fall-back for complete disruptions \citep{Phinney89,Guillochon&RamirezRuiz13,Law-Smith+20}, also in conflict with the shallower decay predicted by $\alpha$-disk models (e.g.~\citealt{Lodato&Rossi11}).  However, in other cases the post-maximum decay is better fit as an exponential (e.g., \citealt{Holoien+16,Blagorodnova+17}) and/or exhibits a flattening at late times (e.g.,~\citealt{Leloudas+16,vanVelzen+19,Wevers+19,Holoien+20}).  Thermal X-rays are detected from a subset of optically-selected TDEs, but the X-ray rise is frequently significantly delayed, by up to hundreds of days, with respect to the optical peak (e.g., \citealt{Gezari+06,Gezari+09,Gezari+17,Kajava+20,Hayasaki&Jonker21,Yao+22,Liu+22}).  Late-time TDE X-ray light curves also frequently decay at a rate more shallow than $\propto t^{-5/3}$ (e.g., \citealt{Holoien+16,Auchettl+17}), which has been attributed to a mismatch between the SMBH accretion rate and fall-back rate due to the viscous time of the disk (e.g., \citealt{Cannizzo+90,Shen&Matzner14,Guillochon&RamirezRuiz15,Auchettl+17}).

In part to address these rapidly growing observational data, numerical (magneto-)hydrodynamical simulations of TDEs have been developed over the past decade (e.g., \citealt{Hayasaki+13,Guillochon&RamirezRuiz13,Shiokawa+15,Bonnerot+16,Hayasaki+16,Sadowski+16,Jiang+16,Steinberg+19,Bonnerot&Lu20,Ryu+21,Bonnerot+21,Curd21,Andalman+22,Steinberg&Stone22}).  Many of these efforts aim to determine what processes lead to debris circularization, with a major focus on physical collisions between outgoing and incoming bound debris streams (e.g., \citealt{Hayasaki+13,Shiokawa+15,Lu&Bonnerot20}), and how such collisions are hastened or delayed by effects such as cooling or general-relativistic precession (e.g., \citealt{Dai+15,Guillochon&RamirezRuiz15,Darbha+19,Andalman+22,Bonnerot&Stone21}).    

These issues bear crucially on the energy source powering TDE flares.  If circularization is significantly delayed (e.g., by many orbits of the most tightly bound debris), then powering the optical luminosities of TDEs requires tapping directly into the limited amount of energy dissipated by stream-stream collisions (e.g., \citealt{Piran+15}).  On the other hand, if even a modest fraction of the bound debris reaches small scales around the SMBH, the resulting accretion power can be sufficient to power the observed UV/optical emission, e.g. via the reprocessing of disk-emitted X-rays by radially-extended material (e.g., \citealt{Metzger&Stone16,Roth+16,Dai+18}), such as bound tidal debris (e.g., \citealt{Guillochon&RamirezRuiz13}), wide-angle unbound outflows generated during the circularization process (e.g., \citealt{Metzger&Stone16,Lu&Bonnerot20}), or accretion disk winds (e.g., \citealt{Strubbe&Quataert09,Miller15,Dai+18,Wevers+19,Uno&Maeda20}).  The geometric beaming of thermal X-rays from the inner accretion flow along the low-density polar regions of the reprocessing structure offers a unification scheme for the optical and X-ray properties of TDEs based on the observer viewing angle (e.g, \citealt{Metzger&Stone16,Dai+18}).  However, while some TDEs exhibit clear evidence for outflows (e.g., \citealt{Miller+15,Alexander+17,Kara+18,Blagorodnova+19,Nicholl+20,Hung+21}), the mass-loss rates required to sustain the large photosphere radii in outflow reprocessing scenarios may in some events be unphysically large (e.g., \citealt{Matsumoto&Piran21}). 

Some observations hint that the peak SMBH accretion rate can be significantly delayed with respect to the optical peak.  While a handful of powerful jetted TDEs exhibit bright non-thermal X-ray and radio emission (e.g., \citealt{Bloom+11,Burrows+11,Levan+11,Zauderer+11}), most TDEs are radio dim (e.g., \citealt{Bower+13,vanVelzen+13,Alexander+20}) excluding powerful off-axis jets (e.g., \citealt{Generozov+17}).  Nevertheless, several TDEs exhibit late-time radio flares, indicating mildly relativistic material ejected from the viscinity of the SMBH, but delayed from the optical peak by several months to years (e.g., \citealt{Horesh+21,Horesh+21b,Perlman+22,Sfaradi+22,Cendes+22}).  A potentially related occurrence is the coincident detection of high-energy neutrinos from three optical TDEs by IceCube \citep{Stein+21,vanVelzen+21b,Reusch+22}, each of which also arrived several months after the optical peak.  State transitions in the accretion flow offer one potential explanation for the delayed onset of jetted accretion activity (e.g., \citealt{Giannios&Metzger11,Tchekhovskoy+14}).  Delayed circularization leading to delayed disk formation (as invoked to explain similarly delayed rises in the X-ray emission; e.g., \citealt{Gezari+17}), offers another.

Even rapid and efficient circularization may not however create a compact disk, at least initially.  \citet{Loeb&Ulmer97} assume the TDE debris forms a spherical radiation-dominated hydrostatic envelope encasing the SMBH. \citet{Coughlin&Begelman14} emphasize that the low angular momenta of TDE debris relative to their binding energy (i.e., ``circularization'' radii $\ll$ ``virial'' radii) endow the circularized structure with properties quite unlike thin Keplerian disks, due to their much larger radial extent and propensity to launch outflows/jets along an extremely narrow polar funnel.  While this structure may briefly manifest as a high-eccentricity disk (e.g., \citealt{Cao+18,Liu+21,Wevers+22}), dissipation within this geometrically thick (e.g., \citealt{Steinberg&Stone22}) confluence of differentially-precessing annuli will likely be strong (e.g., \citealt{Bonnerot+17,Ryu+21}).  And once thermal pressure provides the support against gravity, the bound debris may arguably be modeled most simply as a quasi-spherical ``envelope'' \citep{Loeb&Ulmer97}.  A quasi-spherical emission surface is supported by spectropolarimetry observations of some TDEs (e.g., \citealt{Patra+22}).

Recently, \citet{Steinberg&Stone22} presented three-dimensional radiation hydrodynamical simulations of the tidal disruption of a solar-mass star by a $10^{6}M_{\odot}$ black hole, for the most common (but most computationally challenging) case of a $\beta = 1$ orbit penetration factor.  Unlike the findings or assumptions of most previous works, they find rapid circularization of the debris streams within a short time ($\lesssim 70$ days), comparable to the fall-back time of the most tightly bound debris.  A possible explanation for their result is stronger tidal compression and heating of the streams as they pass through pericenter compared to that found in previous work (e.g., \citealt{Guillochon&RamirezRuiz13,Bonnerot&Lu22a}) due to the inclusion of recombination energy in the assumed equation of state.  \citet{Steinberg&Stone22} further show that radiative diffusion from the extended circularized envelope generates a rising optical light curve with an effective temperature consistent with those of optically-selected TDE flares.  Rapid circularization was also found in general-relativistic hydrodynamical simulations of a similar 1:$10^{6}$ mass-ratio system by \citet{Andalman+22}, in the case of a higher $\beta = 7$ encounter for which rapid stream-stream collisions driven by general relativistic precession play a decisive role in circularization.  

Motivated in part by these recent findings of rapid circularization even across the most commonly sampled regions of TDE parameter space, here we present a model for the long-term evolution and emission from TDE envelopes following their formation.  Though following in spirit \citet{Loeb&Ulmer97}, we make different assumptions and track the time-evolution of the envelope size in light of various cooling and heating processes, including SMBH feedback.  The proposed model of TDE emission as being driven by the thermal evolution of a spherical envelope, while clearly oversimplified in many respects, may nevertheless provide a new view on open questions, such as the timescale and shape of the light curve decay, and the origin of the observed delay between the optical peak and those physical processes (soft X-ray emission, radio flares, fast outflows, neutrino production) instead driven by the innermost SMBH accretion flow.

This paper is organized as followed.  In Sec.~\ref{sec:model} we present the model for the envelope evolution.  In Sec.~\ref{sec:results} we present our results, first focusing on a single fiducial model and dissecting the impact of different physical processes (Sec.~\ref{sec:fiducial}) and then comparing the light curve predictions across a range of star and SMBH properties to TDE observations (Sec.~\ref{sec:range}).  In Sec.~\ref{sec:conclusions} we summarize our findings, expand on some implications, and comment on directions for future work.

\section{Model}
\label{sec:model}

\begin{figure}
    \centering
    \includegraphics[width=0.41\textwidth]{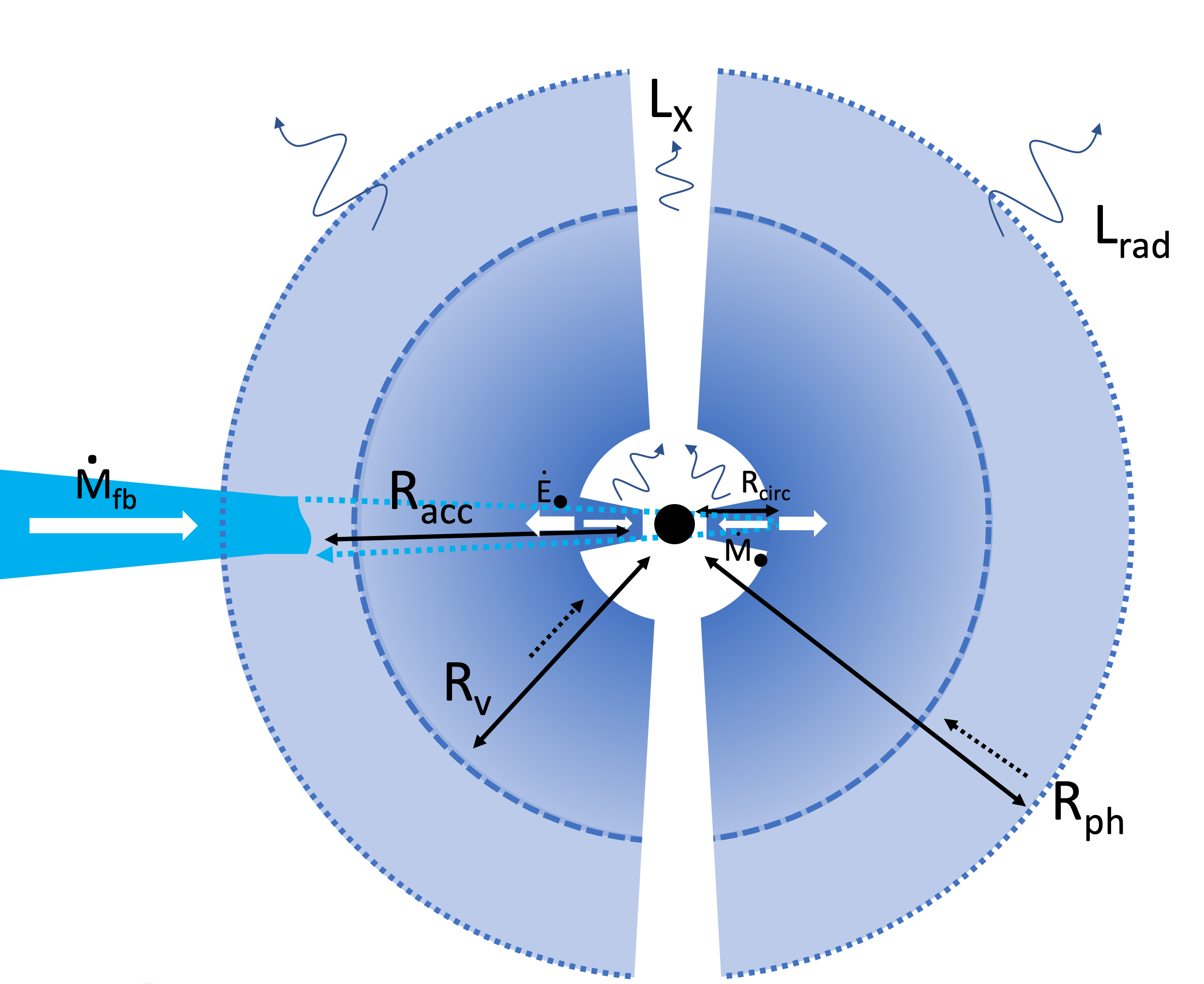}
    \caption{Schematic illustration of the model.  Rapid circularization of the TDE debris forms a quasi-circular pressure-supported envelope around the SMBH of characteristic radius $R_{\rm v}$ and photosphere radius $R_{\rm ph} \sim 10 R_{\rm v}$ which powers the early optical emission.  The envelope luminosity $L_{\rm rad}$ primarily derives from the gravitational energy released from its cooling-driven contraction at close to the Eddington limit, though fall-back accretion (which deposits its energy at a radius $R_{\rm acc} \gtrsim R_{\rm v}$, where the stream dissolves inside the envelope after passing through pericenter) may contribute significantly at early times.  As the envelope cools and contracts, roughly as $R_{\rm v}\propto R_{\rm ph} \propto t^{-1}$ initially, the effective temperature rises $T_{\rm eff} \propto t^{1/2}$ while the optical luminosity $\nu L_{\nu} \propto R_{\rm ph}^{2}T_{\rm eff} \propto t^{-3/2}$ drops.  The SMBH accretion rate rises in tandem, as controlled by the envelope density and viscous time near the circularization radius, potentially powering thermal X-ray and/or jetted activity along the initially narrow polar funnel.  The inner accretion flow also acts as a source of energy to the envelope, which can delay the envelope's contraction in a regulated manner, flattening the late-time optical and X-ray light curve decay.}
    \label{fig:cartoon}
\end{figure}

\begin{deluxetable}{ccc}
\tablecaption{Model Parameters\label{tab:modelparams}}
\tablewidth{700pt}
\tabletypesize{\scriptsize}
\tablehead{
\colhead{Symbol} & \colhead{Description} & 
\colhead{Fiducial Value} 
} 
\startdata
$M_{\star} = m_{\star}M_{\odot}$ & Star mass & $1M_{\odot}$ \\
$M_{\bullet}$ & SMBH mass & $2\times 10^{6}M_{\odot}$ \\
$\beta \equiv R_{\rm t}/R_{\rm p}$ & Orbit penetration factor & 1 \\
$\xi$ & Envelope density power-law (Eq.~\ref{eq:rho}) & 1 \\
$ \zeta$ & Stream penetration factor (Eq.~\ref{eq:zeta}) & 2 \\
$\alpha$ & Viscosity parameter (Eq.~\ref{eq:tacc}) & $10^{-2}$ \\
$H/r$ & Disk aspect ratio (Eq.~\ref{eq:tacc}) & 0.3 \\
$\eta \equiv \dot{E}_{\bullet}/(\dot{M}_{\bullet}c^{2})$ & SMBH feedback efficiency (Eq.~\ref{eq:EdotBH}) & 10$^{-2}$\\
\enddata
\end{deluxetable}

We model the long-term evolution of a quasi-spherical TDE envelope under the assumption of prompt circularization (see Fig.~\ref{fig:cartoon} for a schematic illustration).  We first describe the initial conditions imparted by the tidal disruption process (Sec.~\ref{sec:formation}) and then the details of the envelope evolution (Sec.~\ref{sec:evolution}).

\subsection{Tidal Disruption and Envelope Formation}
\label{sec:formation}

A star of mass $M_{\star} = m_{\star}M_{\odot}$ and radius $R_{\star}$ is tidally disrupted if the pericenter radius of its orbit, $R_{\rm p}$, becomes less than the tidal radius (e.g., \citealt{Hills75})
\begin{eqnarray}
R_{\rm t} &\approx& R_{\star}(M_{\bullet}/M_{\star})^{1/3} \nonumber \\
&\approx& 7\times 10^{12}\,{\rm cm} \,m_{\star}^{7/15}M_{\bullet,6}^{1/3} \approx 47 m_{\star}^{7/15}M_{\bullet,6}^{-2/3}R_{\rm g},
\label{eq:Rt}
\end{eqnarray}
where $M_{\bullet}$ is the SMBH mass, $R_{\rm g} \equiv GM_{\bullet}/c^{2}$, and we assume here and in analytic estimates to follow a mass-radius relationship $R_{\star} \approx m_{\star}^{4/5}R_{\odot}$ appropriate to lower main sequence stars (\citealt{Kippenhahn&Weigert90}).  The orbital penetration factor is defined as $\beta \equiv R_{\rm t}/R_{\rm p} > 1$.

Disruption binds roughly half the star to the SMBH by a specific energy $|E_{\rm t}| = kGM_{\bullet}R_{\star}/R_{\rm t}^{2}$ corresponding roughly to the work done by tidal forces over a distance $\sim R_{\rm t}$.  The most tightly bound matter falls back to the SMBH on the characteristic fall-back timescale set by the period of an orbit with energy $E_{\rm t}$,
\be
 t_{\rm fb} \simeq 58\,{\rm d} \,\,\left(\frac{k}{0.8}\right)^{-3/2}\,M_{\bullet,6}^{1/2}m_{\star}^{1/5},
\label{eq:tfb}
\ee
where the factor $k$ has a weak dependence on the penetration factor $\beta$ (\citealt{Stone+13}; \citealt{Guillochon&RamirezRuiz13}) and hereafter we shall take $k \simeq 0.8$ corresponding to a $\beta = 1$ disruption for a $\gamma = 5/3$ polytropic star.  The resulting rate of mass fall-back at time $t \gg t_{\rm fb}$ is given by\footnote{Partial disruptions$-$which are expected to be volumetrically more common than full disruptions$-$exhibit a steeper fall-back decay, which asymptotes to $\propto t^{-9/8}$ at late times (e.g., \citealt{Guillochon&RamirezRuiz13,Coughlin&Nixon19}).} (e.g., \citealt{Phinney89,Evans&Kochanek89}),
\begin{eqnarray}
&& \dot{M}_{\rm fb} \approx \frac{2M_{\rm acc}}{3 t_{\rm fb}}\left(\frac{t}{t_{\rm fb}}\right)^{-5/3} \nonumber \\
&&\underset{M_{\rm acc} = 0.4M_{\star}}\approx 1.1\times 10^{26}\,\,{\rm g\,s^{-1}}\,M_{\bullet,6}^{-1/2}m_{\star}^{4/5}\left(\frac{t}{t_{\rm fb}}\right)^{-5/3},
\label{eq:Mdotfb}
\end{eqnarray}
where $M_{\rm acc} \simeq 0.4M_{\star}$ is the fraction of the star accreted at $t > t_{\rm fb}$ (the remainder $\sim 0.1M_{\star}$ being accreted during the rise phase at $t < t_{\rm fb}$; this fraction in general depends on the outer density structure of the star). 

The peak Eddington ratio of the fall-back rate,
\be
\frac{\dot{M}_{\rm fb}(t_{\rm fb})}{\dot{M}_{\rm Edd}} \approx 65M_{\bullet,6}^{-3/2}m_{\star}^{4/5},
\label{eq:edd}
\ee
exceeds unity, where $\dot{M}_{\rm Edd} \equiv L_{\rm Edd}/(0.1c^{2})$, $L_{\rm Edd} = 4\pi GM_{\bullet}c/\kappa_{\rm es} \approx 1.4\times 10^{44}M_{\bullet,6}$ erg s$^{-1}$, and $\kappa_{\rm es} \approx 0.35$ cm$^{2}$ g$^{-1}$ is the electron scattering opacity.  

Motivated by recent simulations (e.g., \citealt{Steinberg&Stone22,Andalman+22}) we assume rapid circularization of the initial tidal debris into a quasi-spherical envelope, on a timescale $t_{\rm circ} \sim t_{\rm fb}$.  We further assume the envelope possesses a power-law radial density profile with a characteristic radius $R_{\rm v}$ and a sharp outer edge:
\begin{eqnarray} \label{eq:rho}
\rho = \frac{M_{\rm e}}{4\pi R_{\rm v}^{3}}\frac{(3-\xi)}{(7-2\xi)} 
    \begin{cases}
 \left(\frac{r}{R_{\rm v}}\right)^{-\xi} & 
 r < R_{\rm v}, \\
\exp\left[-(r-R_{\rm v})/R_{\rm v}\right], & r > R_{\rm v},
    \end{cases}       
\end{eqnarray}  
where $\xi \lesssim 3$.  In what follows we take $\xi = 1$, i.e. $\rho \propto r^{-1}$ for $r < R_{\rm v}$;\footnote{\citet{Steinberg&Stone22} found $\rho \propto r^{-1.3}$ out to a break radius $R_{\rm b} \sim 10^{14}$ cm and $\rho \propto r^{-4}$ at $r \gg R_{\rm b}$ (E.~Steinberg, private communication).} however, the qualitative features of the model should be preserved for other choices $1 \lesssim \xi \lesssim 3$ (the ``ZEBRA'' models of \citealt{Coughlin&Begelman14} predict $1/2 < \xi < 3$ for an adiabatic index $\gamma = 4/3$).  Neglecting wind mass-loss or SMBH accretion, the envelope mass grows with time $t > t_{\rm fb}$ as
\begin{eqnarray} &&M_{\rm e}(t > t_{\rm fb}) = M_{\rm e}(t_{\rm fb}) + \int_{t_{\rm fb}}^{t} \dot{M}_{\rm fb}dt' \nonumber \\
&&\underset{M_{\rm acc} = 0.4M_{\star}}\approx 0.1M_{\star} + 0.4 M_{\star}\left[1-\left(\frac{t}{t_{\rm fb}}\right)^{-2/3}\right].
\label{eq:Me0}
\end{eqnarray}

We estimate the characteristic initial radius of the envelope $R_{\rm v}$ (``virial radius'') by equating the energy of the bound stellar debris, $|E_{\rm t}| = k GM_{\bullet}M_{\star}R_{\star}/R_{\rm t}^{2}$, to half of its gravitational binding   energy
$|E_{\rm b}| = 4\pi \int GM_{\bullet}\rho r dr = 4GM_{\bullet}M_{\rm e}/5R_{\rm v}$:
\begin{eqnarray}
R_{\rm v,0} &\approx& \frac{2R_{\star}}{5k}\left(\frac{M_{\bullet}}{M_{\star}}\right)^{2/3}\frac{M_{\rm e}}{M_{\star}} \nonumber \\
&\approx& 6.8\times 10^{13}\,{\rm cm}\,m_{\star}^{2/15} M_{\bullet,6}^{2/3}\left(\frac{M_{\rm e,0}}{0.2M_{\star}}\right),
\label{eq:Rv0}
\end{eqnarray}
where in the second line and hereafter we take $k = 0.8$.  

The envelope is notably much larger than the circularization radius, $R_{\rm circ} = 2R_{\rm t}$,
\begin{eqnarray}
\frac{R_{\rm v,0}}{R_{\rm circ}} &\approx& \frac{\beta}{5k}\left(\frac{M_{\bullet}}{M_{\star}}\right)^{1/3}\frac{M_{\rm e}}{M_{\star}}  \nonumber \\
&\approx& 5.5 \beta m_{\star}^{-1/3}M_{\bullet,6}^{1/3}\left(\frac{M_{\rm e,0}}{0.2M_{\star}}\right).
\end{eqnarray}
Rotational support is thus sub-dominant initially compared to thermal pressure, rendering the envelope quasi-spherical.  The envelope energy we assume is smaller (equivalently, $R_{\rm v}$ is larger) than \citet{Loeb&Ulmer97}, who in adopting a steeper density profile $\rho \propto r^{-3}$ effectively take $R_{\rm v} \sim R_{\rm circ}$ (their Eq.~14), a more tightly bound envelope than imparted by the TDE.    

The characteristic density of the envelope at $r = R_{\rm v}$ is given by
\begin{eqnarray}
\rho_{\rm c} &\equiv& \rho(R_{\rm v}) = \frac{M_{\rm e}}{10\pi R_{\rm v}^{3}} \approx 4\times 10^{-11}{\rm g\,cm^{-3}} m_{\star}^{3/5} \times \nonumber \\  && M_{\bullet,6}^{-2}\left(\frac{M_{\rm e}}{0.2M_{\star}}\right)^{-2}\left(\frac{R_{\rm v}}{R_{\rm v,0}}\right)^{-3}.
\end{eqnarray}
From the virial theorem, the envelope's internal energy $E_{\rm int} = 3 \int P_{\rm rad}4\pi r^2 dr \equiv 4\pi R_{\rm v}^{3}\bar{P}_{\rm rad}$ equals $|E_{\rm t}|$, leading to an estimate of the interior temperature:
\begin{eqnarray}
T_{\rm c} &\equiv& \left(\frac{3\bar{P}_{\rm rad}}{a}\right)^{1/4} = \left(\frac{3GM_{\rm e}M_{\bullet}}{10\pi a R_{\rm v}^{4}}\right)^{1/4} \nonumber \\
&\approx& 4.2\times 10^{5}\,{\rm K}\,m_{\star}^{7/60} \times \nonumber \\
&& M_{\bullet,6}^{-5/12}\left(\frac{M_{\rm e}}{0.2M_{\star}}\right)^{-3/4}\left(\frac{R_{\rm v}}{R_{\rm v,0}}\right)^{-1}.
\end{eqnarray}
Thus, as the envelope contracts and $R_{\rm v}$ decreases, both $\rho_{\rm c}$ and $T_{\rm c}$ will rise.  Radiation pressure $P_{\rm rad} = aT^{4}/3$ dominates over gas pressure $P_{\rm gas} = \rho kT/\mu m_p$ at all times:
\begin{eqnarray}
\frac{P_{\rm rad}}{P_{\rm gas}} \sim  10^{4} \left(\frac{T_{\rm c}}{10^{5}\,{\rm K}}\right)^{3}\left(\frac{\rho_{\rm c}}{10^{-11}\,\rm g\,cm^{-3}}\right)^{-1} \propto R_{\rm v}^{0}
\label{eq:entropy}
\end{eqnarray}
For $\rho \propto r^{-\xi}$, hydrostatic balance $dP_{\rm rad}/dr \propto -GM_{\bullet}\rho/r^{2}$ implies $T \propto r^{-(1+\xi)/4}$ and hence our assumed density profile ($\xi = 1$; Eq.~\ref{eq:rho}) implies $T \propto r^{-1/2}$.  The envelope entropy profile $s \propto T^{3}/\rho \propto r^{(\xi - 3)/4}$ is therefore unstable to convection ($ds/dr < 0$).  While efficient convection may try to drive $s \simeq const$ ($\rho \propto r^{-3}$), for $\xi = 3$ such a configuration would be more tightly gravitationally-bound than permitted by its initial energy.  The envelope structure may therefore try to evolve towards $\xi = 3$ as it cools and contracts, but for simplicity we neglect this possibility and assume $\xi = 1$ at all times. 

\subsection{Envelope Evolution}
\label{sec:evolution}

After forming at time $t_{\rm circ} \approx t_{\rm fb}$, the mass and energy,
\be
|E| = \frac{1}{2}|E_{\rm b}| = \frac{2GM_{\bullet}M_{\rm e}}{5R_{\rm v}},
\ee
(and hence characteristic radius $R_{\rm v}$) of the envelope evolve at later times according to
\be
\frac{dM_{\rm e}}{dt} = \dot{M}_{\rm fb} - \dot{M}_{\bullet} - \dot{M}_{\rm w},
\label{eq:dMdt}
\ee
\be
\frac{d|E|}{dt} = L_{\rm Edd} - \dot{E}_{\bullet} + \dot{E}_{\rm w},
\label{eq:dEdt}
\ee
respectively.  Here $\dot{M}_{\bullet}/\dot{E}_{\bullet}$ accounts for the effects of accretion onto the SMBH, a specific treatment of which is given in Secs.~\ref{sec:accretion}.  The terms $\dot{M}_{\rm w}/\dot{E}_{\rm w}$ allow for mass- and energy-loss in a wind from the envelope given an appropriate prescription relating them to other properties of the system.  Though included for completeness, we neglect outflows hereafter (i.e.~we take $\dot{M}_{\rm w} = \dot{E}_{\rm w} = 0$); we speculate on when this assumption may be violated in Sec.~\ref{sec:conclusions}.  

The luminosity radiated by the envelope can be written
\be
L_{\rm rad} \simeq L_{\rm Edd} + L_{\rm fb} = 4\pi \sigma T_{\rm eff}^{4}R_{\rm ph}^{2},
\label{eq:Lrad}
\ee
where $T_{\rm eff}$ and $R_{\rm ph}$ are the effective temperature and photosphere radius, respectively.\footnote{We neglect the fact that for a scattering-dominated atmosphere, the thermalization surface can be deeper than the photosphere, changing the relationship between effective temperature and the scattering photosphere radius (e.g., \citealt{Lu&Bonnerot20,Steinberg&Stone22}).}  The latter is defined by
\be
\tau = \int_{R_{\rm ph}}^{\infty}\kappa \rho dr = 1.
\ee
For our assumed density profile at $r > R_{\rm v}$ (Eq.~\ref{eq:rho}), 
\be
R_{\rm ph} = R_{\rm v}(1+{\rm ln}\Lambda),
\label{eq:Rphmin}
\ee
where the characteristic optical depth,
\begin{eqnarray}
\Lambda &\equiv& \frac{\kappa M_{\rm e}}{10\pi R_{\rm v}^{2}} \approx 960\,m_{\star}^{-4/15}\times \nonumber \\
&& M_{\bullet,6}^{-4/3}\left(\frac{\kappa}{\kappa_{\rm es}}\right)\left(\frac{M_{\rm e,0}}{0.2M_{\star}}\right)^{-1}\left(\frac{R_{\rm v}}{R_{\rm v,0}}\right)^{-2},
\label{eq:Lambda}
\end{eqnarray}
and we hereafter take $\kappa = \kappa_{\rm es} \simeq 0.35$ cm$^{2}$ g$^{-1}$ (scattering generally dominates other opacity sources given the envelope's high entropy).  

The envelope is in hydrostatic equilibrium and supported by radiation pressure, so its inner layers must radiate close to the SMBH Eddington luminosity, $L_{\rm Edd} \approx 1.4\times 10^{44}\,{\rm erg\,s^{-1}}M_{\bullet,6}$ \citep{Loeb&Ulmer97}, which therefore enters as a loss-term in Eq.~\eqref{eq:dEdt}.  The second term in Eq.~\eqref{eq:Lrad}, 
\be
L_{\rm fb} = \frac{GM_{\bullet}\dot{M}_{\rm fb}}{R_{\rm acc}}   ,
\label{eq:Edotfb}
\ee
accounts for heating of the outer envelope layers by the fall-back stream (specific kinetic energy $v_{\rm ff}^{2}/2 \approx GM_{\bullet}/r$ at radius $r$), where $r \sim R_{\rm acc}$ is the characteristic radius at which the the stream material decelerates and becomes incorporated into the envelope.\footnote{By including $L_{\rm fb}$ in $L_{\rm rad}$ (Eq.~\ref{eq:Lrad}) but not Eq.~\eqref{eq:dEdt}, we have implicitly assumed that $L_{\rm fb}$ is ``instantaneously'' radiated by the envelope.  This is generally a good assumption because the timescale over which $L_{\rm fb}$ evolves $\sim t$ is typically long compared to the Kelvin-Helmholtz time over which the envelope can radiate any deposited energy (see Eq.~\ref{eq:tKHratio}).}

\citet{Steinberg&Stone22} show that the densest portion of the fall-back stream$-$that which remains gravitationally self-bound$-$thickens during compression at pericenter and then dissolves into the envelope on a radial scale $\sim R_{\rm v,0}$ (as a part of the same process giving rise to efficient circularization in the first place).  Motivated thus, we assume the accretion radius scales with the apocenter distance of the fall-back stream,
\be
R_{\rm acc} = \zeta R_{\rm v,0}\left(\frac{t}{t_{\rm fb}}\right)^{2/3},
\label{eq:zeta}
\ee
where the constant $\zeta$ accounts for various uncertainties (e.g., in the radial distribution of the envelope heating and the fraction of the stream able to penetrate the envelope).  We adopt a fiducial value $\zeta = 2$, because only the fraction (e.g.,  $\sim 1/2$, depending on $\beta$; \citealt{Steinberg+19}) of the stream mass confined in the transverse directions by self-gravity (e.g., \citealt{Coughlin+16,Bonnerot+22}) is likely to survive penetration through the envelope to radii $\sim R_{\rm v}.$

\subsubsection{Black Hole Accretion}
\label{sec:accretion}

Insofar as the envelope retains the same specific angular momentum from the time of disruption, rotation will become important on small radial scale $\lesssim R_{\rm circ} = 2R_{\rm p} = 2R_{\rm t}/\beta \ll R_{\rm v}$, thus limiting accretion onto the SMBH based on the viscous time at this radius.  We estimate the accretion rate through the rotationally-supported inner disk region (second term in Eq.~\ref{eq:dMdt}) as 
\be
\dot{M}_{\bullet} \simeq 3\pi \nu \Sigma|_{R_{\rm circ}} \equiv \frac{M_{\rm e}}{t_{\rm acc}}
\label{eq:MdotBH}
\ee
where $\Sigma \simeq \rho(R_{\rm circ}) R_{\rm circ} \simeq M_{\rm e}/(10\pi R_{\rm v}^{2})$, $\nu = \alpha c_{\rm s}^{2}/\Omega = r^{2}\Omega (H/r)^{2}$, $\alpha$ is the \citet{Shakura&Sunyaev73} viscosity parameter, $H/r = c_{\rm s}/(r\Omega)$ is the vertical disk aspect ratio, $c_{\rm s} = (P/\rho)^{1/2}$ the sound speed and $\Omega = (GM_{\bullet}/r^{3})^{1/2}$ the Keplerian orbital frequency.  An ``accretion'' timescale for the envelope can thus be defined,
\begin{eqnarray}
&t_{\rm acc} &= \frac{10}{3\alpha}\frac{R_{\rm v}^{2}}{(GM_{\bullet}R_{\rm circ})^{1/2}}\left(\frac{H}{r}\right)^{-2} \approx 12.8\,{\rm yr}\,\,\frac{\beta^{1/2}}{\alpha_{-2}} \times \nonumber \\
&& m_{\star}^{1/30}M_{\bullet,6}^{2/3}\left(\frac{H}{0.3r}\right)^{-2}\left(\frac{M_{\rm e,0}}{0.2M_{\star}}\right)^{2}\left(\frac{R_{\rm v}}{R_{\rm v,0}}\right)^{2},
\label{eq:tacc}
\end{eqnarray}
where $\alpha_{-2} \equiv \alpha/(10^{-2}$) and we are motivated to consider a geometrically thick disk $H/r \gtrsim 0.3$, consistent with the near-Eddington accretion rates of interest \citep{Abramowicz+88,Shen&Matzner14}.  Though initially much longer than other timescales in the problem, $t_{\rm acc}$ will shorten $\propto R_{\rm v}^{2}$ as the envelope cools and contracts.

Finally, the third term in Eq.~\eqref{eq:dEdt} accounts for energy released by accretion onto the SMBH which is transferred outwards to the envelope through radiation or convection.  We assume the feedback luminosity scales with the accretion rate at $R_{\rm circ}$, 
\begin{eqnarray}
\label{eq:EdotBH}
&& \dot{E}_{\bullet} = \eta \dot{M}_{\bullet}c^{2}  ,
\end{eqnarray}
where the dimensonless efficiency $\eta \lesssim 0.1$ encapsulates a number of uncertain factors related to efficiency of reprocessing by the envelope of radiation/outflows/jets from the inner disk, including the potential for disk-wind mass-loss from the (potentially super-Eddington) accretion flow between $R_{\rm circ}$ and $R_{\rm isco}$ (e.g., \citealt{Blandford&Begelman99}).  A canonical minimum value, set by the condition $\dot{E}_{\bullet} = GM_{\bullet}\dot{M}_{\bullet}/R_{\rm circ}$, is 
\be \eta_{\rm min} = \frac{R_{\rm g}}{R_{\rm circ}} \approx 10^{-2}\beta m_{\star}^{-7/15}M_{\bullet,6}^{2/3}.
\label{eq:etamin}
\ee

\subsection{Summary of the Model}
\label{sec:modelsummary}

As summarized in Table \ref{tab:modelparams}, a given model is defined primarily by the masses of the star $M_{\star}$ and SMBH $M_{\bullet}$.  Secondary variables, whose values we shall typically fix within a fiducial value or range, include: orbital penetration factor $\beta = 1$; density radial profile power-law $\xi = 1$; fall-back heating efficiency $\zeta =2$; viscosity $\alpha = 0.01$, aspect ratio $H/r = 0.3$, and feedback efficiency $\eta \sim \eta_{\rm min} \sim 10^{-2}-10^{-1}$ of the inner accretion disk.  Starting the calculation at time $t = t_{\rm fb}$, when the envelope mass $M_{\rm e}(t_0) = 0.1M_{\star}$ (Eq.~\ref{eq:Me0}) and radius $R_{\rm v} = R_{\rm v,0}$ (Eq.~\ref{eq:Rv0}), we solve Eqs.~\eqref{eq:dMdt}, \eqref{eq:dEdt} for the evolution of $M_{\rm e}$, radius $R_{\rm v}$, photosphere radius $R_{\rm ph}$, and effective temperature $T_{\rm eff}$ of the envelope, as well as the SMBH accretion rate $\dot{M}_{\bullet}$.
We evolve the system until the envelope mass reaches zero, though the assumptions of the model may break down before this, once the radius contracts to $R_{\rm v} \lesssim R_{\rm circ} = 2R_{\rm t}/\beta$ violating the assumption of negligible rotational support.  Given the bolometric luminosity $L_{\rm rad}$ and $T_{\rm eff}$ we calculate the luminosity $\nu L_{\nu}$ at a given optical waveband $\nu$ assuming blackbody emission (e.g., we neglect the distinction between the scattering photosphere and frequency-dependent thermalization surface, which can lead to an underestimate of the luminosity on the Rayleigh-Jeans tail; e.g., \citealt{Lu&Bonnerot20}).  

\section{Results}
\label{sec:results}

We begin in Sec.~\ref{sec:fiducial} by showing results for a fiducial model with $M_{\star} = M_{\odot}$, $M_{\bullet} = 2\times 10^{6}M_{\odot}$, $\beta = 1$.  Rather than including all of the physics into the model at once, we begin (Sec.~\ref{sec:KH}) by artificially neglecting the effects arising from accretion onto the SMBH (i.e., we assume $\dot{M}_{\bullet} = \dot{E}_{\bullet} = 0$) and walking through some analytic arguments which reproduce the results.
Then we move on to models which include SMBH accretion, first just a mass-loss term for the envelope (Sec.~\ref{sec:massloss}) and then finally including also SMBH energy feedback (Sec.~\ref{sec:BHheating}).  Finally, Sec.~\ref{sec:range} presents the full-model optical/X-ray light curves for a range of star and SMBH properties.

\subsection{Fiducial Model}
\label{sec:fiducial}

\subsubsection{Pure Cooling (Kelvin-Helmholtz Contraction)}
\label{sec:KH}

\begin{figure}
    \centering
    \includegraphics[width=0.45\textwidth]{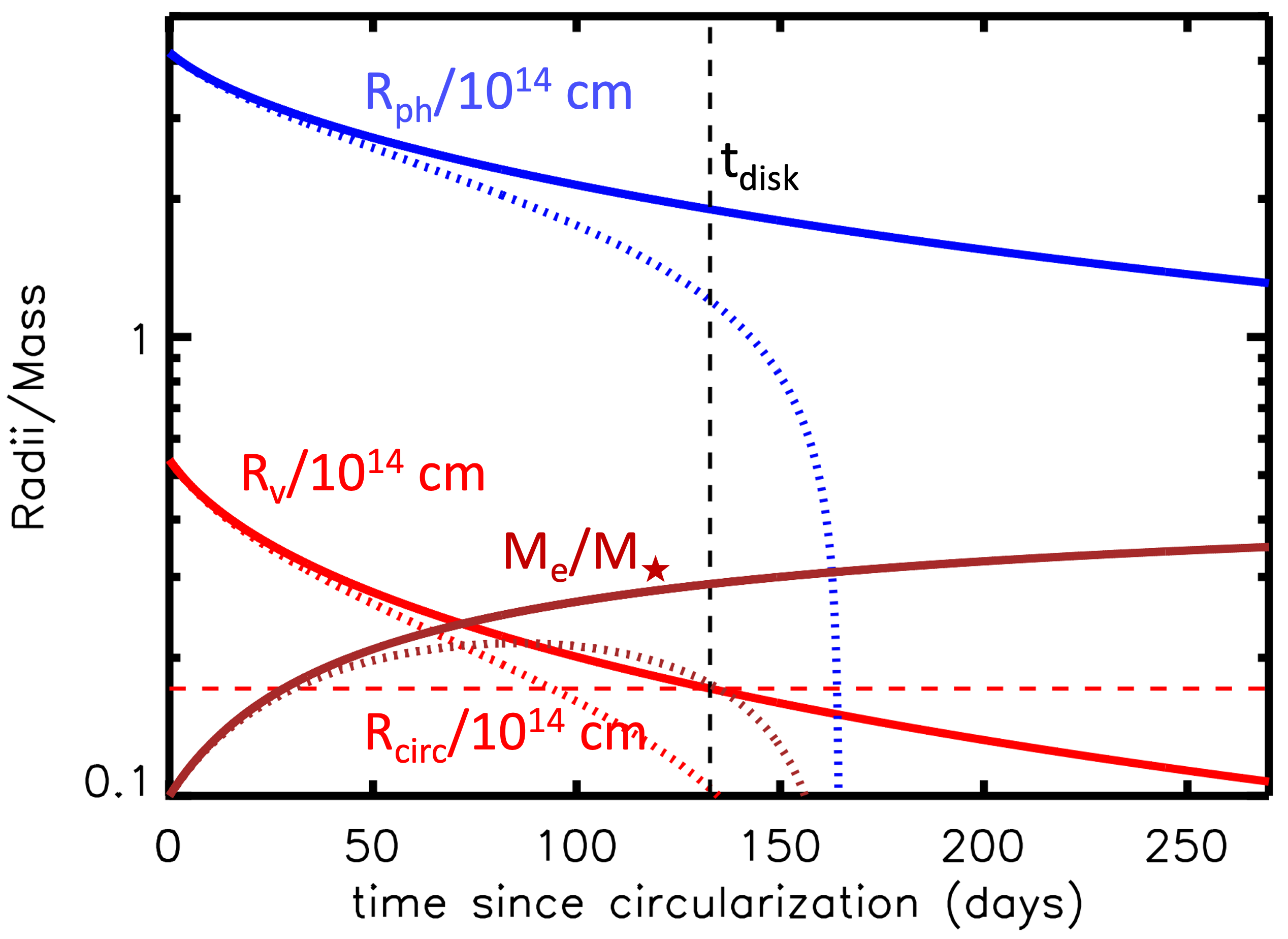}
    \includegraphics[width=0.50\textwidth]{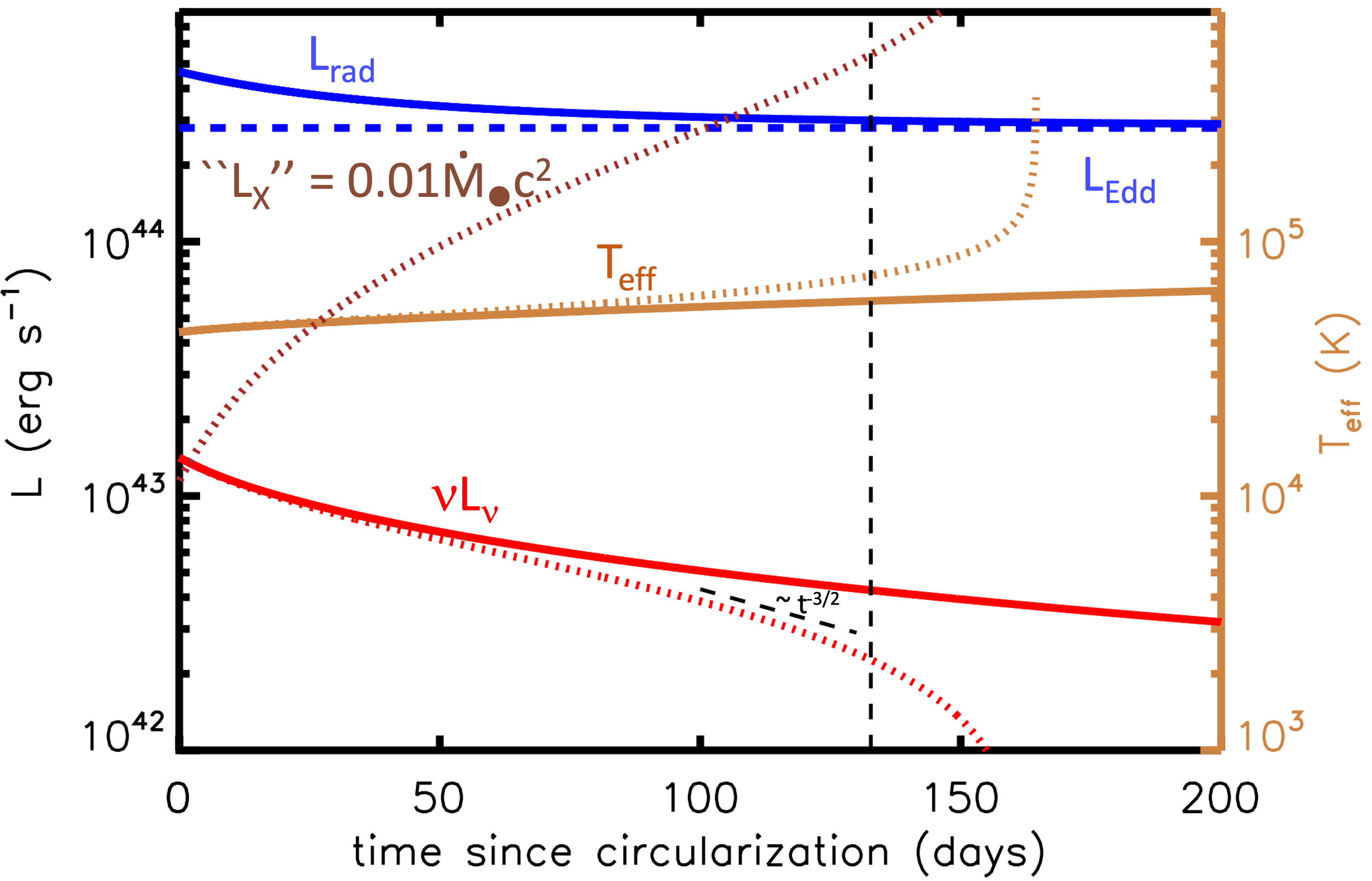}
    \caption{Evolution of TDE cooling envelope model with $M_{\star} = M_{\odot}$, $M_{\bullet} = 2\times 10^{6}M_{\odot}$, $\beta = 1$, shown as a function of time since the circularization/envelope-assembly time $t = t_{\rm fb}$.  Solid lines show a model neglecting mass-accretion onto the SMBH (i.e., $\dot{M}_{\bullet} = \dot{E}_{\bullet} = 0$), while dotted lines show the effects of including SMBH accretion mass-loss (Eq.~\ref{eq:MdotBH} for $\alpha = 0.01$, $H/r = 0.3$) but not energy feedback.  The top panel shows the characteristic envelope radius $R_{\rm v}$, mass $M_{\rm e}/M_{\star}$, and photosphere radius $R_{\rm ph}$, while the bottom panel shows the envelope luminosity $L_{\rm rad}$, optical luminosity $\nu L_{\nu}$ at frequency $\nu = 6\times 10^{14}$ Hz (g-band), effective temperature $T_{\rm eff}$, and $L_{\rm X} \equiv 0.01\dot{M}_{\bullet}c^{2}$, taken as a proxy for the X-ray luminosity from the inner disk (potentially observable only through a narrow polar region).  The envelope evolution concludes roughly once $R_{\rm v}$ decreases to $R_{\rm circ}$ (horizontal dashed line), as occurs roughly at the times $t_{\rm disk}$ (Eq.~\ref{eq:tdisk}) and $t_{\rm acc}^{\star}$ (Eq.~\ref{eq:taccstar}), respectively, in the two models.  A dashed black line illustrates $\propto t^{-3/2}$ decay.  }
    \label{fig:KH}
\end{figure}

Solid lines in Fig.~\ref{fig:KH} show results for the envelope evolution, neglecting accretion or feedback onto the SMBH.  The envelope radius $R_{\rm v}$ and photosphere radius $R_{\rm ph}$ begin large, but gradually decay with time, with $R_{\rm v}$ reaching $R_{\rm circ}$ by around day 130 measured with respect to the envelope assembly (time $t_{\rm circ} \sim t_{\rm fb}$ after the disruption).  Likewise, while the bolometric luminosity is at or slightly above the Eddington luminosity of the SMBH at all times, the optical decays from its initial value $\nu L_{\nu} \gtrsim 10^{43}$ erg s$^{-1}$ roughly $\propto t^{-3/2}$, as the effective temperature rises.  These results can largely be understood analytically.

Neglecting fall-back heating, Eq.~\eqref{eq:dEdt} becomes,
\be
\frac{2GM_{\bullet}M_{\rm e}}{5R_{\rm v}^{2}}\frac{dR_{\rm v}}{dt} = L_{\rm Edd}(M_{\bullet}).
\ee
Further approximating the envelope mass $M_{\rm e}$ as a constant (in reality, $M_{\rm e}$ grows gradually from $0.1M_{\odot}$ to $0.5M_{\star}$), we obtain
\be
\frac{2GM_{\bullet}M_{\rm e}}{5}\left(\frac{1}{R_{\rm v}}-\frac{1}{R_{\rm v,0}}\right) = L_{\rm Edd}t.
\ee
The radius contracts as,
\be
\frac{R_{\rm v}}{R_{\rm v,0}} = \left(1+\frac{5L_{\rm Edd}tR_{\rm v,0}}{2GM_{\bullet}M_{\rm env}}\right)^{-1} = \left(1 + \frac{t}{t_{\rm KH,0}}\right)^{-1},
\label{eq:Rv}
\ee
where 
\be
t_{\rm KH} \equiv \frac{2GM_{\bullet}M_{\rm e}}{5L_{\rm Edd}R_{\rm v}} \simeq 24\,{\rm d}\,m_{\star}^{13/15}M_{\bullet,6}^{-2/3}\left(\frac{R_{\rm v}}{R_{\rm v,0}}\right)^{-1}
\label{eq:tKH}
\ee
is the ``Kelvin-Helmholtz'' time and $t_{\rm KH,0} = t_{\rm KH}(R_{\rm v} = R_{\rm v,0})$ defines its initial value.\footnote{The thermal timescale $t_{\rm KH}$ also equals the photon diffusion time through the envelope, $t_{\rm diff} \sim \Lambda(R_{\rm v}/c),$ where $\Lambda$ is the characteristic optical depth (Eq.~\ref{eq:Lambda}).} Making use of Eq.~\eqref{eq:tfb}, we see that
\be
\frac{t_{\rm KH} }{t} \approx  0.41\,m_{\star}^{2/3}M_{\bullet,6}^{-7/6}\left(\frac{R_{\rm v}}{R_{\rm v,0}}\right)^{-1}\left(\frac{t}{t_{\rm fb}}\right)^{-1}.
\label{eq:tKHratio}
\ee
The fact that $t_{\rm KH}/t \ll 1$ at times $t \gtrsim t_{\rm fb} \sim t_{\rm circ}$ implies that (1) thermal equilibrium can be established on the timescale the envelope is being assembled and will remain so at later times; (2) if the assembly process itself is not rapid (taking place over a timescale $\gtrsim 0.1-1 t_{\rm fb}$, depending on the SMBH mass), then the light curve properties near peak light will depend on the assembly history and hence may not be captured by our model, which assumes instantaneous assembly (we return to this point in Sec.~\ref{sec:range}).

Fig.~\ref{fig:KH} shows that the envelope luminosity (Eq.~\ref{eq:Lrad}) roughly obeys $L_{\rm rad} \simeq L_{\rm Edd}$ with the fall-back luminosity $L_{\rm fb}$ (Eq.~\ref{eq:Edotfb}) boosting this value only moderately at early times.  Indeed, from Eqs.~\eqref{eq:Mdotfb}, \eqref{eq:zeta}, \eqref{eq:Rv} we obtain,
\begin{eqnarray}
&&\frac{L_{\rm fb}}{L_{\rm Edd}} \simeq \frac{GM_{\bullet}\dot{M}_{\rm fb}}{R_{\rm acc}L_{\rm Edd}} \approx \frac{GM_{\bullet}\dot{M}_{\rm fb}}{\zeta R_{\rm v} L_{\rm Edd}} \nonumber \\
&&\underset{t \gg t_{\rm KH,0}}\approx 0.7\left(\frac{\zeta}{2}\right)^{-1} \left(\frac{M_{\rm e,0}}{0.2M_{\star}}\right)^{-1}\left(\frac{t}{t_{\rm fb}}\right)^{-4/3},
\label{eq:Lfbratio}
\end{eqnarray}
a result which is notably independent of $m_{\star}$ and $M_{\bullet}.$  Fallback accretion thus contributes at an order-unity level to the envelope luminosity at early times $t \sim t_{\rm fb}$, but becomes comparatively less important with time relative to the passive envelope cooling.
 
Approximating $L_{\rm rad} \simeq L_{\rm Edd}$ and using Eqs.~\eqref{eq:Lrad}, \eqref{eq:Rphmin}, \eqref{eq:Rv}, the effective temperature evolves according to
\begin{eqnarray}
&& T_{\rm eff} \simeq \left(\frac{L_{\rm Edd}}{4\pi \sigma R_{\rm ph}^{2}}\right)^{1/4} 
\simeq  4.5\times 10^{4}\,{\rm K}\,m_{\star}^{-1/15}\times \nonumber \\
&& M_{\bullet,6}^{-1/12}\left(\frac{M_{\rm e,0}}{0.2M_{\odot}}\right)^{-1/2}\left(\frac{\rm ln \Lambda}{10}\right)^{-1/4}\left(1+\frac{t}{t_{\rm KH,0}}\right)^{1/2}.
\label{eq:Teff}
\end{eqnarray}
The predicted gradual rise in $T_{\rm eff} \propto t^{1/2}$ is consistent with that shown in Fig.~\ref{fig:KH} and similar to that of observed TDE UV/optical flares (e.g., \citealt{vanVelzen+21}; their Fig.~5).  

The optical luminosity, in the Rayleigh-Jeans approximation $h \nu \ll kT_{\rm eff}$, evolves as
\begin{eqnarray}
\nu L_{\nu} &\simeq& \frac{8\pi^{2}}{c^{2}}\nu^{3} k T_{\rm eff} R_{\rm ph}^{2} \nonumber \\
&\simeq& 5.4\times 10^{43}{\rm erg\,s^{-1}\,} \left(\frac{\nu}{6\times 10^{14}\,{\rm Hz}}\right)^{3}m_{\star}^{1/5} \times \nonumber \\
&& M_{\bullet,6}^{5/4}\left(\frac{M_{\rm e,0}}{0.2M_{\star}}\right)^{3/2}\left(\frac{\rm ln\Lambda}{10}\right)^{7/4}\left(1+\frac{t}{t_{\rm KH,0}}\right)^{-3/2},
\label{eq:nuLnu}
\end{eqnarray}
again in reasonable agreement with Fig.~\ref{fig:KH}.  The predicted $\propto t^{-3/2}$ power-law decay is (coincidentally) similar to the mass fall-back rate, $\dot{M}_{\rm fb} \propto t^{-5/3}$ (Eq.~\ref{eq:Mdotfb}).

Envelope contraction as we have modeled it will continue until rotational support becomes important, as occurs once $R_{\rm v}$ decreases to $R_{\rm circ} = 2 R_{\rm t}$.  Again neglecting accretion onto the SMBH, this timescale for the envelope to transform into a disk, can be estimated using Eq.~\eqref{eq:Rv} (in the $t \gg t_{\rm KH,0}$ limit):
\begin{eqnarray}
& t_{\rm disk}& \simeq \frac{R_{\rm v,0}}{2\beta R_{\rm t}}t_{\rm KH,0} \simeq \frac{1}{5\beta}\frac{GM_{\bullet}M_{\rm e,0}}{L_{\rm Edd}R_{\rm t}} \nonumber \\
&&\approx  121\,{\rm d} \, m_{\star}^{8/15}M_{\bullet,6}^{-1/3},
\label{eq:tdisk}
\end{eqnarray}
i.e. typically several months to a year, in agreement with where $R_{\rm v}$ crosses $R_{\rm circ}$ in Fig.~\ref{fig:KH}.

\subsubsection{Mass-Loss from SMBH Accretion}
\label{sec:massloss}

A dotted line Fig.~\ref{fig:KH} shows an otherwise identical model to that presented in the previous, but which now includes envelope mass-loss due to SMBH accretion (Eq.~\ref{eq:MdotBH}, assuming $\alpha = 10^{-2}$ and $H/r = 0.3$), yet still neglects any feedback heating from the accretion.  At early times the solution is similar to that neglecting accretion, until around day 100 when the envelope mass reaches a maximum and begins to decrease.  This in turn drives $R_{\rm v}$ and $R_{\rm ph}$ to decrease, and thus $T_{\rm eff}$ to increase and $\nu L_{\nu}$ to drop, at a faster rate than they would otherwise without accretion.

We can estimate the time required for the envelope to be fully accreted by setting $t = t_{\rm acc}$ (Eq.~\ref{eq:tacc}) with $R_{\rm v} \underset{t \gg t_{\rm KH,0}}= R_{\rm v,0}(t_{\rm KH,0}/t)$ (Eq.~\ref{eq:Rv}), which gives
\begin{eqnarray}
&&t_{\rm acc}^{\star} \simeq \frac{0.72}{\alpha^{1/3}}\left(\frac{H}{r}\right)^{-2/3}\frac{(GM_{\bullet})^{1/2}M_{\rm e,0}^{2/3}}{L_{\rm Edd}^{2/3}R_{\rm t}^{1/6}} \approx  146\,{\rm d}\,\alpha_{-2}^{-1/3}\times \nonumber \\
&& \left(\frac{H}{0.3r}\right)^{-2/3}m_{\star}^{53/90}M_{\bullet,6}^{-2/9}\left(\frac{M_{\rm e,0}}{0.2M_{\star}}\right)^{2/3},
\label{eq:taccstar}
\end{eqnarray} 
in rough agreement with where $M_{\rm e}$ begins to fall rapidly in Fig.~\ref{fig:KH}.  Depending on the value of $\alpha(H/r)^{2}$, $t_{\rm acc}^{\star}$ can be larger or smaller than $t_{\rm disk}$, the maximum disk formation time absent accretion (Eq.~\ref{eq:tdisk}).

Accretion onto the central SMBH can in principle power X-ray emission, which may begin to escape along what may be a narrow accretion funnel (e.g., \citealt{Kara+18,Dai+18}) to a greater and greater fraction of external observers as the envelope becomes more disk-like ($R_{\rm v} \rightarrow R_{\rm circ}$).  A brown dotted line in Fig.~\ref{fig:KH} shows an estimate of this proxy X-ray luminosity,
\be
L_{\rm X} \equiv 10^{-2}\dot{M}_{\bullet}c^{2},
\label{eq:Lx}
\ee
where the prefactor $10^{-2}$ is an estimate of the radiative efficiency of super-Eddington accretion disks (e.g., \citealt{Sadowski&Narayan16}).  Although the normalization of the X-ray power is clearly uncertain, and the observed luminosity be highly inclination-dependent (e.g., \citealt{Dai+18}), the key feature of note is the {\it delayed rise} of the X-ray light curve relative to the optical peak.  Though such delayed X-ray rises are observed in some TDEs, they have frequently been attributed to inefficient or delayed circularization (e.g., \citealt{Shiokawa+15,Gezari+17}).  The physics here is instead {\it delayed cooling} and envelope contraction, which leads to accelerating growth in the SMBH accretion rate $\dot{M}_{\bullet} \propto t_{\rm acc}^{-1} \propto R_{\rm v}^{-2}$ (Eq.~\ref{eq:tacc}).

\subsubsection{Feedback from SMBH Heating}
\label{sec:BHheating}

\begin{figure}
    \centering
    \includegraphics[width=0.42\textwidth]{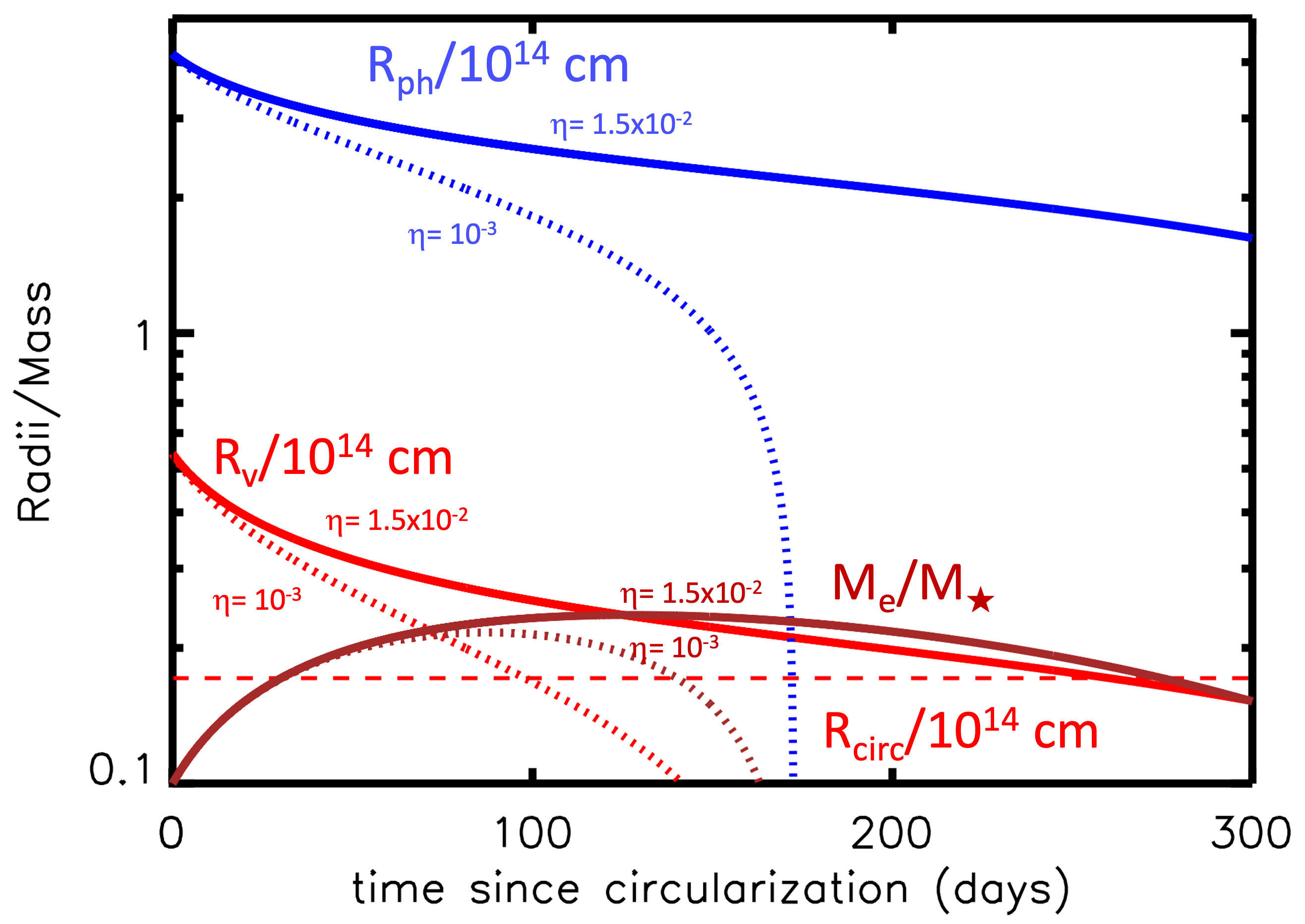}
    \includegraphics[width=0.5\textwidth]{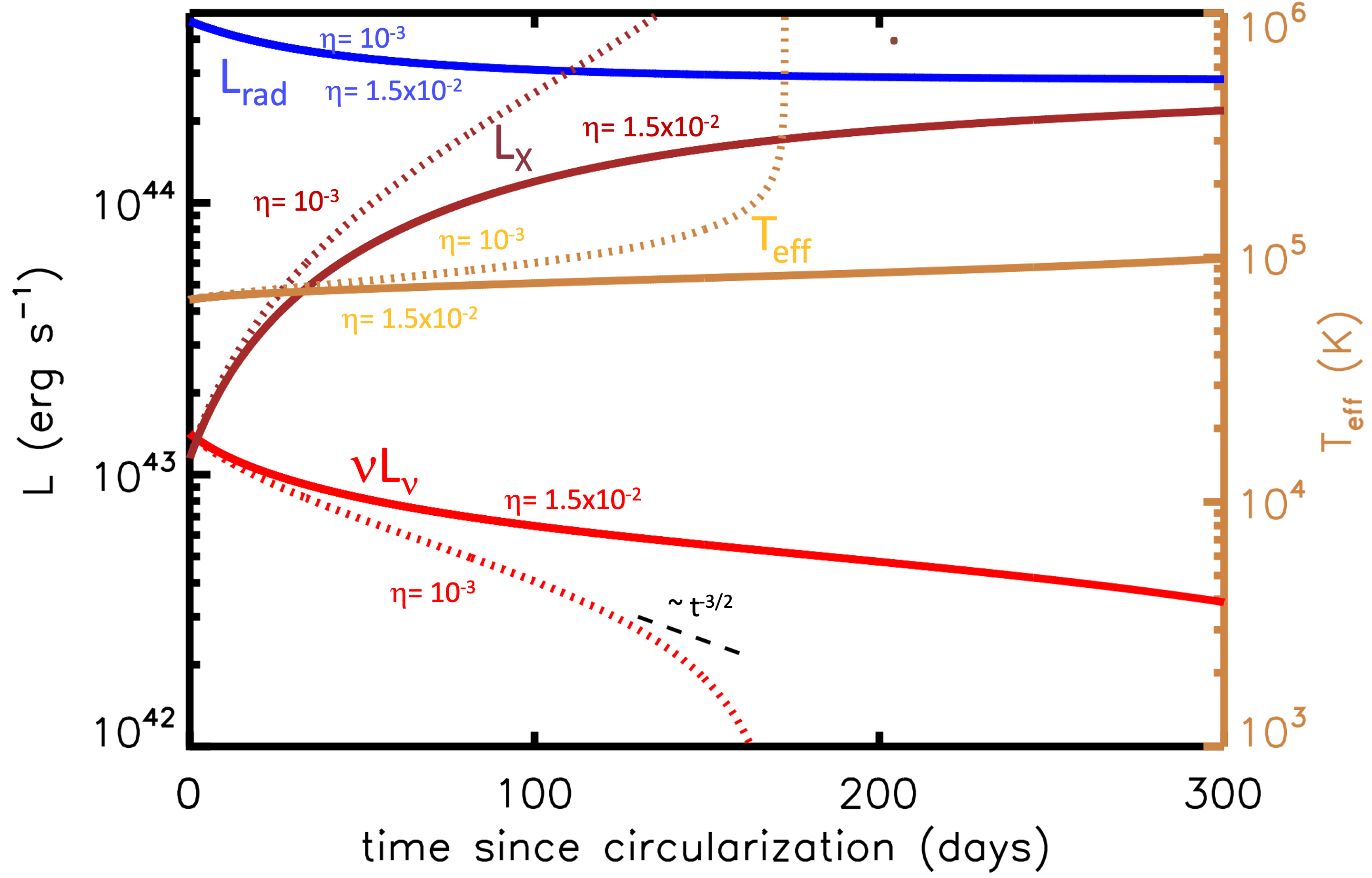}
    \caption{Models calculated for the same parameters as in Fig.~\ref{fig:eta}, but now including the effects of SMBH feedback on the envelope structure for two different values of $\eta = 1.5\times 10^{-2}$ (solid lines) and $\eta = 10^{-3}$ (dotted lines).  Powerful feedback (large $\eta$) acts to slow the rate of envelope cooling and accretion, flattening the late-time $\nu L_{\nu}$ optical light curve decay.  A dashed black line shows $\propto t^{-3/2}$ decay, as roughly expected absent efficient feedback. }
    \label{fig:eta}
\end{figure}

Finally, in Fig.~\ref{fig:eta} we show the effects of adding SMBH accretion heating to the envelope evolution (Eq.~\ref{eq:EdotBH}), by comparing a model with a low feedback efficiency $\eta = 10^{-3}$ (dotted line) to one with higher efficiency $\eta = 1.5\times 10^{-2}$ (solid line).  The $\eta = 10^{-3}$ model follows a similar evolution to models excluding feedback altogether (Fig.~\ref{fig:KH}).  However, the $\eta = 1.5\times 10^{-2}$ model differs markedly, exhibiting a much more gradual decline in the rate of envelope contraction and optical luminosity.  In effect, the energy provided by SMBH accretion keeps the envelope ``puffed up'' for longer, which in turn slows the SMBH accretion rate.

This regulated state, in which SMBH feedback reaches a balance with the radiated luminosity $L_{\rm rad} \simeq L_{\rm Edd}$, can be expressed as a condition on the SMBH accretion rate (see also \citealt{Loeb&Ulmer97})
\be
\dot{E}_{\bullet} \simeq L_{\rm Edd} \Rightarrow \dot{M}_{\bullet} = \frac{L_{\rm Edd}}{\eta c^{2}}.
\label{eq:MdotBHreg}
\ee
Equating this with Eq.~\eqref{eq:MdotBH}, the corresponding envelope radius in the SMBH-regulated state is given by
\begin{eqnarray}
&&R_{\rm v}^{\bullet} \simeq \left(\frac{3\eta \alpha c^{2}}{10}\left(\frac{H}{r}\right)^{2}\frac{M_{\rm e}}{L_{\rm Edd}}\right)^{1/2}\left(GM_{\bullet}R_{\rm circ}\right)^{1/4}\nonumber \\
&&\approx 1.7\times 10^{13}{\rm cm}\,\frac{\eta_{-2}^{1/2}\alpha_{-2}^{1/2}}{\beta^{1/4}}\left(\frac{H}{0.3r}\right)m_{\star}^{37/60}M_{\bullet,6}^{-1/6}\left(\frac{M_{\rm e}}{0.2M_{\star}}\right)^{1/2}
\end{eqnarray}
The timescale for the envelope to be completely accreted at the regulated rate (Eq.~\ref{eq:MdotBHreg}) is thus given by,
\be
t_{\rm acc}^{\bullet} = \frac{M_{\rm e}}{\dot{M}_{\bullet}} = \frac{\eta M_{\rm e} c^{2}}{L_{\rm Edd}} \simeq 300\,{\rm d}\,\eta_{-2}m_{\star}M_{\bullet,6}^{-1}\left(\frac{M_{\rm e}}{0.2M_{\star}}\right)
\label{eq:tacc_regulated}
\ee
From when $R_{\rm v}$ decreases from its initial value to $R_{\rm v}^{\bullet}$, until the time the envelope is accreted $t \sim t_{\rm acc}^{\bullet}$, the envelope radius and optical luminosity will exhibit a flat plateau-like time-evolution (suggestive of the late-time behavior of some TDE light curves; e.g., \citealt{Leloudas+16,vanVelzen+19,Wevers+19}).

The regulated plateau state is only be achieved if $t_{\rm acc}^{\bullet} > t_{\rm acc}^{\star}$, as occurs for sufficiently high accretion feedback efficiency,
\begin{eqnarray}
\eta &>& \eta_{\rm crit} \approx 5\times 10^{-3}\alpha_{-2}^{-1/3}\left(\frac{H}{0.3r}\right)^{-2/3} \times \nonumber \\
&& m_{\star}^{-37/90}M_{\bullet,6}^{7/9}\left(\frac{M_{\rm e}}{0.2M_{\star}}\right)^{-1/3},
\end{eqnarray}
consistent with the large difference in the light curve duration between the $\eta = 1.5\times 10^{-2}$ and $\eta = 10^{-3}$ models in Fig.~\ref{fig:eta}.  For fiducial values of $\alpha$ and $H/r$, $\eta_{\rm crit}$ is comparable to $\eta_{\rm min}$ (Eq.~\ref{eq:etamin}); this suggests that events both with and without a self-regulated plateau phase could occur amongst the TDE population depending on the precise system parameters.

\subsection{Dependence on SMBH/Star Properties}
\label{sec:range}

\begin{figure}
    \centering
    \includegraphics[width=0.42\textwidth]{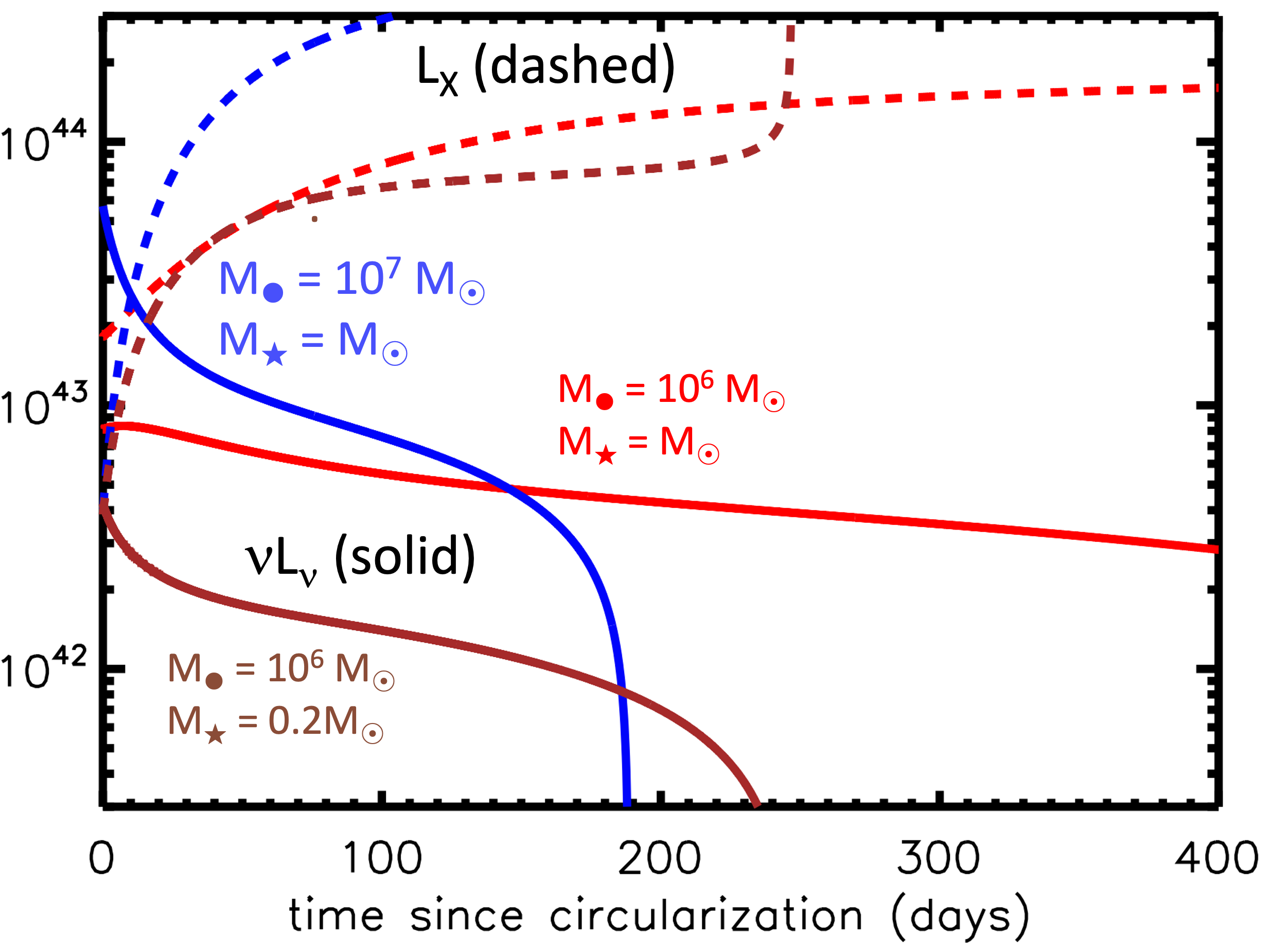}
    \caption{Optical-band light curves $\nu L_{\nu}$ at $\nu = 6\times 10^{14}$ Hz and proxy X-ray light curves $L_{\rm X} = 10^{-2}\dot{M}_{\bullet}c^{2}$ for a series of models which adopt fiducial parameters ($\beta = 1$; $\alpha = 10^{-2}$; $H/r = 0.3$; $\zeta = 2$) but varying the mass of the star and SMBH as marked.  We assuming a SMBH feedback parameter $\eta = \eta_{\rm min}(M_{\star},M_{\bullet}).$ (Eq.~\ref{eq:etamin}).}
    \label{fig:cases}
\end{figure}

Figure \ref{fig:cases} shows the optical $\nu L_\nu$ and proxy X-ray $L_{\rm X} = 10^{-2}\dot{M}_{\bullet}c^{2}$ light curves for a series of models which adopt fiducial parameters but varying the star and SMBH mass as marked and fixing $\eta = \eta_{\rm min}(M_{\star},M_{\bullet})$ (Eq.~\ref{eq:etamin}).  The optical luminosity is higher for more massive stars or SMBHs, consistent with Eq.~\eqref{eq:nuLnu}.  TDEs of lower-mass stars and/or by higher-mass SMBH also tend to produce faster decaying optical light curves (and correspondingly faster rising proxy X-ray light curves), as expected because of their shorter envelope cool times (Eq.~\ref{eq:tKH}).  Since $\eta_{\rm min} \gtrsim \eta_{\rm crit}$ for the assumed values of $\{\beta,\alpha,H/r\}$, the total light curve durations (400, 150, 250 d, respectively) are boosted moderately by SMBH feedback, roughly in accord with the scalings $t_{\rm acc}^{\bullet} \propto m_{\star}^{8/15}M_{\bullet}^{-1/3}$ (Eq.~\ref{eq:tacc_regulated}).  

\citet[hereafter \citetalias{vanVelzen+21}]{vanVelzen+21} analyze the optical light curve properties of a sample of 17 TDEs detected by the Zwicky Transient Facility, exploring internal correlations between the light curve properties (e.g., blackbody luminosity $L_{\rm rad}$, blackbody[our photosphere] radius $R_{\rm ph}$, effective temperature $T_{\rm eff}$, rise time $t_{\rm rise}$, decay time $t_{\rm decay}$) and with the host galaxy stellar mass $M_{\rm gal}$ (a rough proxy for the SMBH mass given $M_{\rm gal}$-$M_{\bullet}$ and related correlations; e.g. \citealt{Magorrian+98}).  We briefly describe comment on our model's predictions in terms of their findings.

Albeit with large scatter, \citetalias{vanVelzen+21} find evidence for a positive correlation between the blackbody luminosity and host galaxy mass.  This supports more luminous TDEs arising from higher mass SMBH, consistent with the Eddington-limited luminosity $L_{\rm rad} \propto L_{\rm Edd}$ of a hydrostatic envelope (as predicted in our model given the likely sub-dominant role played by fall-back accretion luminosity throughout the bulk of the light curve; Eq.~\ref{eq:Lfbratio}).

\citetalias{vanVelzen+21} also find that the flare rise-time is correlated with peak luminosity and anti-correlated with the photosphere radius.  Insofar as $t_{\rm rise}$ is set by the timescale of envelope assembly $\sim t_{\rm circ}$ (e.g., \citealt{Steinberg&Stone22}), it could be expected to scale with the fall-back time (Eq.~\ref{eq:tfb}), i.e.
\be
t_{\rm rise} \sim t_{\rm circ} \sim t_{\rm fb} \propto m_{\star}^{1/5}M_{\bullet}^{1/2},
\label{eq:trise}
\ee
leading a correlation between $t_{\rm rise} \propto M_{\bullet}^{1/2} \propto L_{\rm rad}^{1/2}$.

The situation regarding the light curve decay-time is more complicated.  At face value our model predicts that the initial decay-time should scale with the initial envelope cooling time (Eq.~\ref{eq:tKH}), i.e.
\be
t_{\rm decay} \propto t_{\rm KH,0} \propto m_{\star}^{2/3}M_{\bullet}^{-7/6},
\ee
thus predicting a negative correlation between $t_{\rm decay}$ and $M_{\bullet}$ and hence between $t_{\rm decay}$ and $M_{\rm gal}$, contradicting the positive correlation found by \citetalias{vanVelzen+21}.  However, as already mentioned, because $t_{\rm KH,0}/t_{\rm fb} < 1$ (Eq.~\ref{eq:tKHratio}) the light curve shape near peak may be influenced by the envelope assembly processes if the process is not sufficiently abrupt (assembly duration $\ll t_{\rm fb}$).  A significant contribution from fall-back heating at $t \sim t_{\rm fb}$ (Eq.~\ref{eq:Lfbratio}) could also imprint some $t_{\rm fb}-$dependence into the early light curve decay.  Finally, feedback heating from the SMBH also acts to flatten the light curve and increase $t_{\rm decay}$, and may become more efficient for higher SMBH masses because the circularization radius is typically deeper within the gravitational potential well.

The strongest correlation found by \citetalias{vanVelzen+21} is between $L_{\rm pk}/R_{\rm ph}$ and $t_{\rm rise}.$  Taking $L_{\rm pk} \propto M_{\bullet}$ and $R_{\rm ph} \propto R_{\rm v} \propto  m_{\star}^{2/15}M_{\bullet}^{2/3}$, and eliminating the SMBH mass, our model would predict
\be
\frac{L_{\rm pk}}{R_{\rm ph}} \propto m_{\star}^{-2/15}M_{\bullet}^{1/3} \propto m_{\star}^{-4/15}t_{\rm rise}^{2/3}
\ee
While the weak dependence on stellar-mass is encouraging for generating a tight correlation, the scaling with $t_{\rm rise}$ is somewhat too shallow compared to the data (\citetalias{vanVelzen+21}; their Fig. 9).

\section{Conclusions}
\label{sec:conclusions}

We have presented a model for TDE light curves, which starts from the assumption that circularization of the most tightly bound stellar debris is prompt (i.e., occurs on the fall-back time of the most tightly bound debris), resulting in the creation of a quasi-spherical pressure-supported envelope surrounding the SMBH \citep{Loeb&Ulmer97} with a characteristic size much larger than the circularization radius defined by the angular momentum of the original orbit \citep{Coughlin&Begelman14}.  This assumption is motivated by recent hydrodynamical simulations which find prompt circularization and rising optical emission consistent with observed early phases of TDE flares, even for the most common ``garden-variety'' $\beta = 1$ disruptions \citep{Steinberg&Stone22}.  Our model builds on earlier works starting with \citet{Loeb&Ulmer97}, but focuses on predicting the long-term evolution of the envelope, and its accretion rate onto the SMBH, under the influence of different sources/sinks of mass and energy, in a flexible and simple to implement and interpret format.

The proposed ``cooling envelope'' model accounts for a variety of TDE observations, including (1) large photosphere radii and correspondingly high optical luminosities; (2) optical light curve decay, driven largely by passive cooling of the envelope, which roughly follows a power-law $\nu L_{\nu} \propto t^{-3/2}$, coincidentally similar to the canonical $\propto t^{-5/3}$ rate of fall-back decline; (3) potential at late times for a shallower or plateau-shaped light curve decay, due to self-regulated energy input from SMBH accretion; (4) gradually rising effective temperature $T_{\rm eff} \propto t^{1/2}$ as the envelope contracts; (5) delay in the peak of the SMBH accretion rate, and hence of thermal X-ray (for opportunely oriented viewers) or jetted emissions, relative to the time of optical peak by up to several hundred days.  This delay is notably driven by envelope cooling (either acting in isolation, or temporarily offset by SMBH accretion heating), rather than requiring a delay in the circularization process.

An $\sim$Eddington-limited hydrostatic envelope scenario appears broadly consistent with correlations between TDE light curve properties and host galaxy (proxy SMBH) mass \citep{vanVelzen+21}.  On the other hand, the model is challenged to explain the observed positive correlation between proxy SMBH mass and optical decay-time assuming the latter tracks the Kelvin-Helmholtz time $t_{\rm KH,0} \propto M_{\bullet}^{-7/6}$ (Eq.~\ref{eq:tKH}); however, the shortness of $t_{\rm KH}$ relative to the fall-back time $t_{\rm fb} \propto M_{\bullet}^{1/2}$ (Eq.~\ref{eq:tKHratio}) suggests the light curve shape near peak will be sensitive to the envelope assembly process and early-time fall-back heating (Eq.~\ref{eq:Lfbratio}), possibly mixing some $t_{\rm fb}-$dependence into the decay time.

The end of our calculation, and thus of the most optically-luminous phase, is defined by when the envelope contracts to the circularization radius, after which point rotational effects dominate and the disk structure should better resemble the pure $\alpha$-disk models originally envisioned (e.g., \citealt{Rees88,Ulmer99,Lodato&Rossi11}).  The properties of the remaining envelope at this transition may then define the initial conditions for a viscous disk evolution phase (e.g., \citealt{Cannizzo+90,Shen&Matzner14}), which can power longer lasting UV/X-ray emission (e.g., \citealt{Auchettl+17,vanVelzen+19,Jonker+20}).  The timescale of this transition depends on whether the envelope contraction is limited by radiative cooling or accretion, and whether SMBH feedback slows the latter, but in general can roughly be written as:
\be
t_{\rm life} = {\rm min}\left[t_{\rm disk},{\rm max}[t_{\rm acc}^{\star},t_{\rm acc}^{\bullet}]\right],
\ee
where $t_{\rm disk}, t_{\rm acc}^{\star}, t_{\rm acc}^{\bullet}$ are given in Eqs.~\eqref{eq:tdisk}, \eqref{eq:taccstar}, \eqref{eq:tacc_regulated}, respectively.

The prediction of a cooling-induced time delay of several months or longer between the peak of the optical light curve and the SMBH accretion rate, may also bear on other puzzling TDE observations.  One of these is the discovery of late-time radio flares or rebrightenings (e.g., \citealt{Horesh+21,Horesh+21b,Perlman+22,Sfaradi+22,Cendes+22}), which may indicate the delayed ejection of mildly relativistic material from the immediate viscinity of the SMBH several months to years after the optical peak.  We speculate these could arise from jets or winds from the inner accretion disk that suddenly become more powerful as the SMBH accretion rate rises rapidly near the termination of the envelope cooling-contraction phase.  Shocks driven by such outflows into the surrounding wind/envelope material could in principle accelerate relativistic ions, generating a source of high-energy gamma-rays and neutrinos \citep{Senno+17,Lunardini&Winter17,Guepin+18,Fang+20,Murase+20}, perhaps explaining the significant observed delay between the high-energy neutrino detections from a growing sample of TDE and the optical light curve maximum \citep{Stein+21,vanVelzen+21b,Reusch+22}.

Although our model is constructed to allow for the effects of winds or outflows from the envelope on its evolution (the sink terms $\dot{M}_{\rm w}, \dot{E}_{\rm w}$ in Eqs.~\ref{eq:dMdt}, \ref{eq:dEdt}), we have neglected this possibility for simplicity in this work.  Strong outflows could occur from the envelope if energy is deposited below its surface at a highly super-Eddington rate \citep{Quataert+16}.  We speculate this may occur at two phases in the TDE: (1) at early times, when the envelope is being assembled and $R_{\rm acc} \lesssim R_{\rm v}$ is small and hence $\dot{E}_{\rm fb} \gg L_{\rm Edd}$ is possible (Eq.~\ref{eq:Lfbratio}); (2) at late times as $R_{\rm v} \rightarrow R_{\rm acc}$ and the SMBH accretion rate is quickly rising to high values, on a timescale faster than the envelope can radiate the received energy.

\acknowledgements
I am grateful to Elad Steinberg and Nicholas Stone for helpful comments and for sharing an early draft of their manuscript, which imparted momentum to this work.


\bibliographystyle{aasjournal}

\end{document}